# Design of Phase Locked Loop in 180 nm Technology

Priyam Kumar, Akshada Khele, Aditee C. Joshi*

ajoshi@unipune.ac.in ; aditee04@gmail.com

**ABSTRACT**

The presented paper introduces a design for a phase-locked loop (PLL) that is utilized in frequency synthesis and modulation-demodulation within communication systems and in VLSI applications. The CMOS PLL is designed using 180 nm Fabrication Technology on Cadence Virtuoso Tool with a supply voltage of 1.8 V. The performance is evaluated through simulations and measurements, which demonstrate its ability to track and lock onto the input frequency.

The PLL is a frequency synthesizer implemented to generate 2.4 GHz frequency. The input reference clock from a crystal oscillator is 150 MHz square wave. Negative feedback is given by divide-by-16 frequency divider, ensuring the phase and frequency synchronization between the divided signal and the reference signal. The design has essential components such as a phase frequency detector, charge pump, loop filter, current-starved voltage-controlled oscillator (CSVCO), and frequency divider. Through their collaborative operation, the system generates an output frequency that is 16 times the input frequency.

The centre frequency of the 3-stage CSVCO is **3.208 GHz** at **900 mV** input voltage. With an input voltage ranging from **0.4 V to 1.8 V**, the VCO offers a tuning range that spans from **1.066 GHz to 3.731 GHz**. PLL demonstrates a lock-in range spanning from **70.4 MHz to 173 MHz**, with an output frequency range of **1.12 GHz to 2.78 GHz**. It achieves a lock time of **260.03 ns** and consumes a maximum power of **5.15 mW** at **2.4 GHz**.

## I. INTRODUCTION

Phase locking, a concept introduced in the 1930s, quickly gained significant popularity in the fields of electronics and communication [1]. Primary goal of the PLL is to achieve synchronization between the output oscillator signal and a reference signal. In the context of a phase difference, the term "leading phase" denotes a wave that occurs before another wave of the same frequency, while the term "lagging phase" refers to waves that occur after another wave of the same frequency.

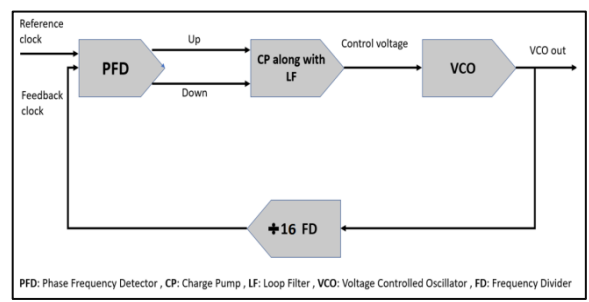

**Fig. 1. Block Diagram of PLL**

A phase-locked loop reduces phase discrepancy between the output and input frequencies. When the phase difference between these signals reaches zero, the system is considered "locked." The PLL achieves this locking behavior through the utilization of negative feedback.

**A. PLL working principle**

**1. Phase Frequency Detector (PFD):** By comparing the phase and frequency of the reference signal with the phase of the feedback signal, the PFD generates an error signal that is proportionate to the phase difference. PFD has two outputs UP and DOWN, which are generated according to which input is leading.

**2. Charge Pump:** The charge pump takes the error signal and produces a current that regulates the voltage-controlled oscillator. During the high state of the UP signal, a positive current flows through the circuit, resulting in charging of capacitor. When the DOWN signal transitions to a high state, a negative current is induced in the circuit, leading to a decrease in the control voltage, thus discharging of capacitor.

**3. Loop Filter:** This block converts the current from the charge pump into a voltage signal, which represents the average output of the phase detector. This voltage signal is then provided as an input to the VCO. This design utilizes a second-order low-pass filter, which eliminates noise and high-frequency components from the charge pump output, which increases stability.

**4. Voltage Controlled Oscillator:** It is an electronic device that produces oscillations whose frequency is determined by an applied input voltage that is controlled. By varying the applied control voltage, the frequency of oscillation can be adjusted. The control input voltage governs the output frequency of the VCO.

**5. Frequency Divider (FD):** Frequency divider is a module that reduces the frequency of a signal. The input signal to the phase detector is derived from the VCO output, passing through a frequency divider that scales down the output frequency to match the reference frequency.

## II. CIRCUIT IMPLEMENTATION

The design and evaluation of various blocks of phase locked loop are explained in the section.

**A. Phase Frequency Detector:**

PFD can be understood from the Fig. 2. and 3 The clock inputs of the flip-flops are linked to inputs A and B. When initially both QA and QB are at a logic low state (QA = QB = 0), if input A transitions to a logic high state, the output QA will rise accordingly. Similarly, When a rising transition occurs on input B, the subsequent action for another flip-flop is that QB goes high. NAND and Inverter are combined to form AND gate which is used to resets both the flip flops. In other words, both QA and QB experience a simultaneous peak for a brief duration, the difference in average values between them effectively represents the input phase or frequency difference.

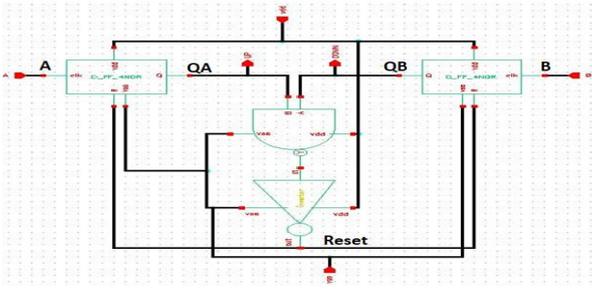

**Fig. 2. Implementation of PFD**

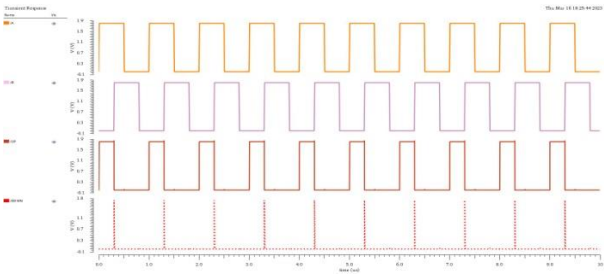

**Fig. 3 Output of PFD**

Each flip flop can be implemented as shown in Fig. 4 where two RS latches are interconnected in a cross-coupled manner. Latch 1 is triggered by the rising edges of the CLK signal, while Latch 2 responds to the rising edges of the Reset signal [1]. This design has advantage that it require less number of transistor and generating the required waveform for phase and frequency detection. It also overcome dead zone problem as it has delay element in the reset path which is AND gate [2].

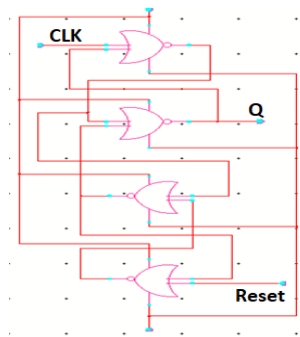

**Fig. 4 Implementation of D-FF**

Fig. 5 illustrates PFD Layout.

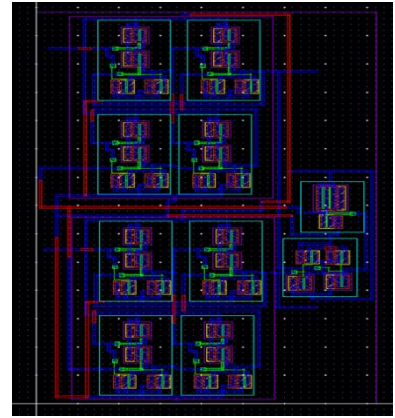

**Fig. 5 Layout of PFD**

### B. Charge Pump along with Loop Filter:

A charge pump circuit comprises a current source and two inputs from a phase frequency detector (PFD) with latches, enabling control over the current flow into and out of the filter. The filter, typically a low-pass filter, is integrated with the charge pump after its implementation. The design of the filter relies on second-order derivatives, and the values of resistors and capacitances are determined using these derivatives [3]. Fig. 6 illustrate the circuit, where the PFD outputs are connected to the charge pump circuit containing the current source implementation, and the loop filter is integrated to the charge pump, with its resistance and capacitances calculated accordingly.

The equation determines the output current of the charge pump [3],

$$I_{PDI} = K_{PDI} * \Delta\Phi \quad (1)$$

Where,

$K_{PDI} = I_{pump}/2*\Pi (A/radian)$  (2)

$\Delta\Phi = \Phi_{in} - \Phi_{ref}$  (3)

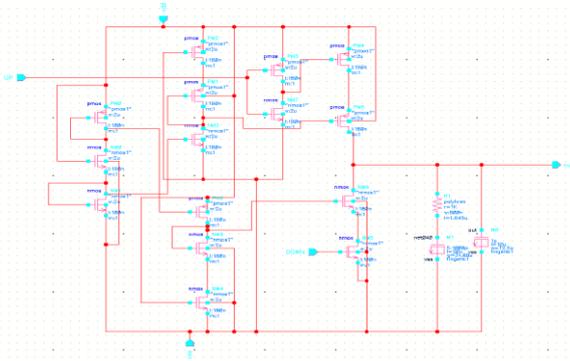

**Fig. 6 Schematic of CP along with LF**

When the loop is in lock, the phase difference (ΔΦ) becomes zero. The input voltage to the VCO is determined by this condition.

$V_{invco} = K_f \times I_{PDI}$  (4)

where $K_f$ represents loop filter gain. Fig. 7 displays the simulation result of the Charge Pump output in response to the generation of the UP signal.

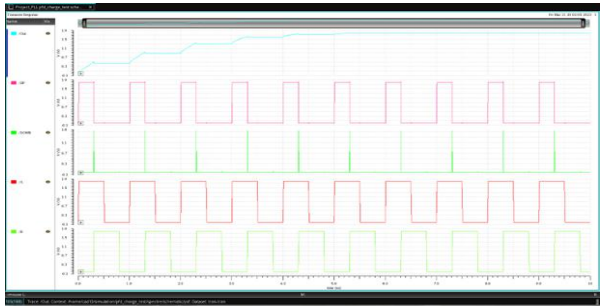

**Fig. 7 Simulation of PFD and CP along with LF**

Fig. 8 displays the layout of the charge pump and loop filter integration. The area measured for CP with LF is (34.37 $\mu$m* 45.12 $\mu$m) 1550.7 $\mu$m$^2$.

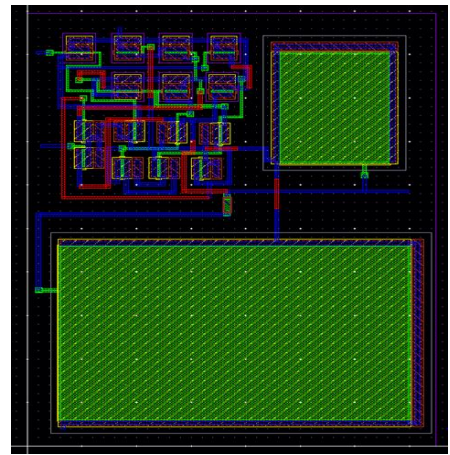

**Fig.8 Layout of CP with LF**

**C. Voltage Controlled Oscillator:**

In this work, a voltage-controlled oscillator (VCO) based on the current-starved principle is implemented. Starving meaning limiting the current. By limiting the current flowing through the inverter, it is possible to keep the oscillation frequency constant. The design of the current-starved VCO incorporates a ring oscillator, and its functionality closely resembles that of a typical ring oscillator[4].A ring oscillator is formed by cascading an odd number of inverters in a series configuration.The output of the final inverter is connected back to the input of the initial inverter in a feedback configuration.For our design number of inverter stages is fixed with Three. At a supply voltage of 900mV, the VCO is designed to achieve a center frequency of 3.208 GHz.

The Schematic of 3 stage VCO is shown in Fig. 9

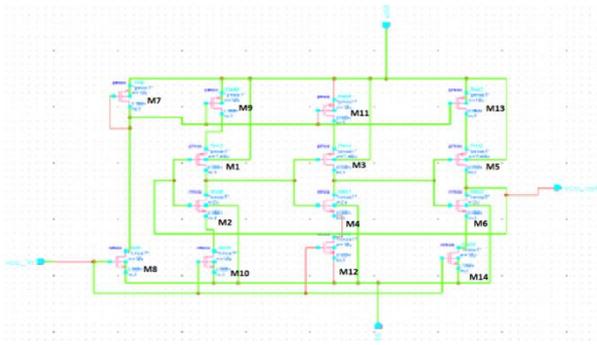

**Fig.9 Schematic of 3 stage CSVCO**

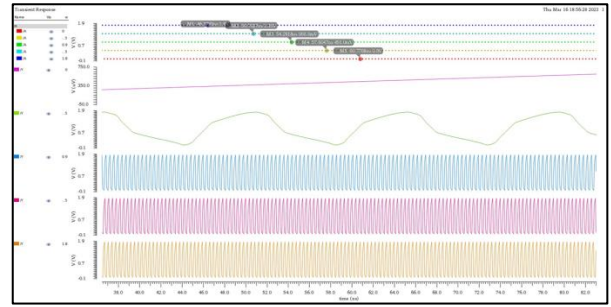

**Fig.10 Simulation result of CSVCO**

The M1 and M2 MOSFETs forms an inverter, whereas M9 and M10 function as current sources. The M9 and M10 MOSFETs restricts the amount of current that can flow to the M1 and M2 inverter, effectively starving it for current. Both M7 and M8 MOSFETs have equal drain currents, which are determined by the input control voltage. Each stage of the inverter and current source mirrors the currents flowing through M7 and M8. The gate of M7 is linked to the upper PMOS transistors, All the low NMOS transistors have their gates connected to the source voltage. The ability to tune the oscillation frequency over a wide range is a benefit of this configuration, achieved by adjusting the value of the control voltage[5],[6]. Moreover, [7] shows linearized CSVCO by keeping resister in series with M8.

$K_{vco} = \frac{f_{max} - f_{min}}{V_{max} - V_{min}}$ (Hz/V)

(5)

Hence, Kvco comes out as 1.265 (GHz/V).

Fig. 10 presents the simulation results of the VCO at various voltages.

The performance analysis for the CSVCO for generating the different frequencies with different control voltages are shown below in the graph in Fig 11.

The table below shows different parameters of CSVCO according to the design configuration.

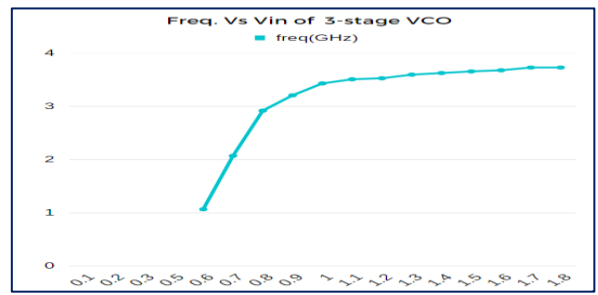

**Fig. 11. Freq. Vs Vin of CSVCO**

**Table I. Design Parameters of CSVCO**

| Sr. no | Parameters | Observation |
|---|---|---|
| 1. | Technology | 180nm |
| 2. | Supply voltage | 1.8V |
| 3. | No. of transistors | 14 |
| 4. | Power Dissipation | 1.60 mW |
| 5. | Center Frequency | 3.208 GHz |
| 6. | Kvco | 1.265 GHz/V |
| 7. | Tunning range | 1.066GHz-3.731GHz |
| 8. | Tunning Voltage | 0.4-1.8V |

Layout of CSVCO in Fig. 12 with area of 464.69 $\mu m^2$.

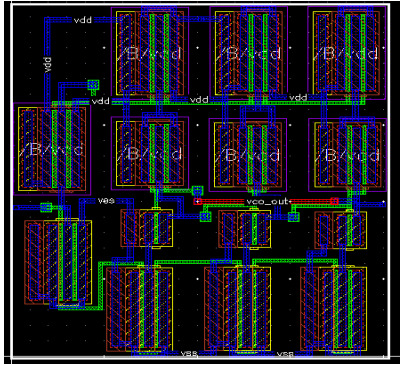

**Fig. 12. Layout of CSVCO**

### D. Frequency Divider:

Through the frequency divider circuit, the output of the VCO is looped back to phase frequency detector input. The frequency divider circuit reduces the frequency of the output signal coming from VCO. This circuit is essentially a D flip-flop (DFF) with its output connected back to its input, forming a feedback loop in Fig. 13. A frequency divider with a division ratio of 16 is employed in this design. Four D ffs are cascaded to form divide by 16 divider. Now there are two kind of FFs which are used for frequency division.

1) Strong- ARM latched based flip flops and
2) TSPC based flip flops [8].

TSPC based FFs has lesser power consumption in general as compared to strong -ARM latched based FFs. It is also quite fast and as there are lesser number of transistors it operates at higher frequency. Thus, TSPC based frequency division method is used in this design [9].

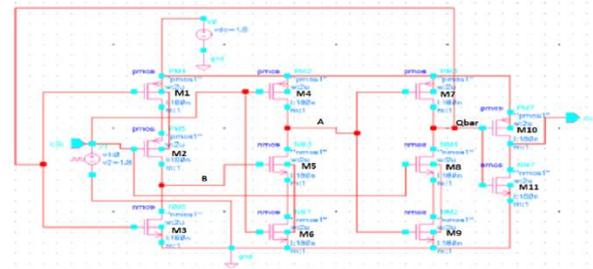

**Fig.13 Schematic of Divide-by-2 D-ff using TSPC Logic**

When clock is zero, at that time if D is 1 then M1 will turned off. Transistor M3 will be ON thus B is 0. So the particular Signal A is irrespective of the input. At clock 0, A is 1. The M7 Transistor is Cut-off. Thus the Q bar value will be there as it is, as M8 will also Off. When Clock goes high, M2 will be Off. B remains at 0. M4 and M6 will turned on. As a result A remains high. Transistor M7 gets off. And m8 and m9 turned on. Thus Q bar signal goes low.

Power consumption of Divide by 2 Counter at different frequencies is given in chart below.

**Table II. Power consumption of Divide-by-2 TSPC D-ff**

| VCO Output Frequency | Counter output Frequency | Power ($\mu W$) |
|---|---|---|
| 1 MHz | 500.9 kHz | 0.24 |
| 10 MHz | 5.02 MHz | 2.3 |
| 100 MHz | 49.99 MHz | 9.022 |
| 1 GHz | 500.4 MHz | 94.27 |
| 2 GHz | 1.0013 GHz | 188.8 |
| 3 GHz | 1.50 GHz | 362.1 |

Cascading four divide by 2 D-ff will give divide by 16 frequency divider as depicted below in fig. 14 and 15.

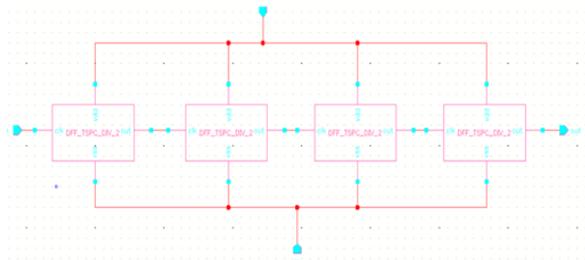

**Fig.14 Schematic of divide by 16 FD**

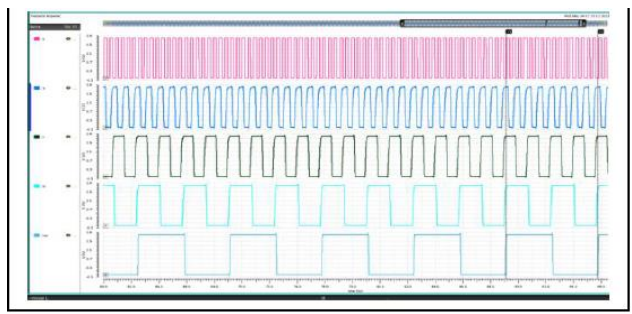

**Fig. 15 Waveform of Divide-by-16 Divider at 2.4 GHz input**

The area measured for CP with LF is (34.37 $\mu$m* 45.12 $\mu$m) 1550.7 $\mu m^2$ as seen in Fig.16.

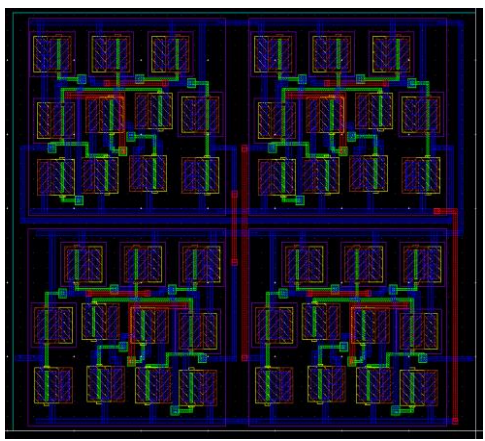

**Fig. 16 Layout of Divide-by-16 Divider**

### III. INTEGRATION OF ALL BLOCKS

After designing and simulating all the blocks such as PFD, CP, LF, VCO and divide by 16 divider, It is important to integrate all these blocks to form PLL. Following is the schematic of the PLL in Fig. 17.

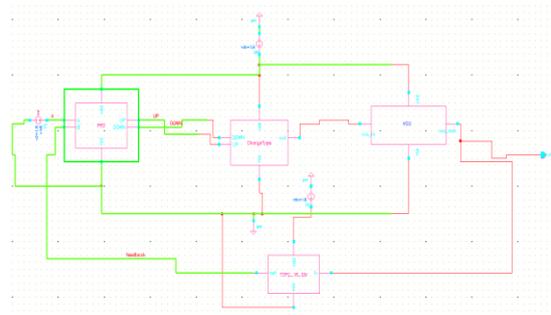

**Fig.17 Schematic of PLL**

Fig. 18 exhibits the simulation results of the PLL in a locked state, operating at reference frequency of 150 MHz.

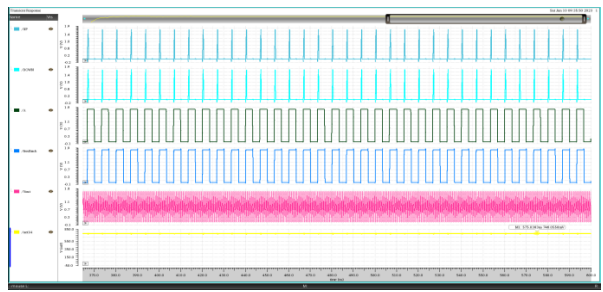

**Fig.18 Output of PLL**

### IV. CONCLUSION

The PLL circuit is created and tested using the Cadence Virtuoso 180nm CMOS Technology, with a supply voltage of 1.8V.The output frequency of the CSVCO is varying from **1.06 GHz to 3.731 GHz**, depending on the input voltage of the VCO. The PLL consumes a maximum power of **5.15 mW**. Currently, the PLL is functioning correctly and generating an output frequency of **2.4 GHz**. The design of PLL is made using **111** transistors. So, the area of chip is reduced and also the power consumption.

**Table III. Simulation Result of PLL**

| Parameters | Observation |
|---|---|
| Technology | 180 nm |
| Power consumption | **5.15 mW** |
| Lock Time | 260.03 ns for 2.4 GHz |
| Power Supply | 1.8 V |
| Lock Range | 1.066 GHz to 3.731 GHz |
| No. of Transistors | 111 |